\documentstyle[11pt,newpasp,epsf]{article}
\begin{document}

\title{How to Form (Twin) Globular Clusters?}
\author{Christian Theis}
\affil{Institut f.\ Theoretische Physik und Astrophysik, Universit\"at Kiel,
24098 Kiel, Germany, email: theis@astrophysik.uni-kiel.de}


\begin{abstract}
  Though it is generally assumed that massive molecular clouds are the 
  progenitors of globular clusters, their detailed formation mechanism 
  is still unclear. Standard scenarios based on the collapse of a 
  smooth matter distribution suffer from strong requirements with respect
  to cluster formation time scale, binding energy and star formation 
  efficiency. An alternative model assuming cluster formation due to
  the recollapse of a supernova-induced, fragmented shell can relax
  these difficulties.

    In this paper the final collapse stages of the different scenarios are 
  compared by 
  N-body simulations for shells and spheres. It is shown that fragmentation 
  is much more pronounced for shells. Taking a galactic tidal field into account
  shells preferably 
  form twin (or multiple) systems, whereas spheres end up as single 
  clusters. The twins are characterized by identical metallicities,
  and stellar mass functions; some of them show counter-rotating cores.
  Their orbital evolution can result in both, a final merger or well
  separated twins sharing a common galactic orbit.
\end{abstract}



\section{Introduction}

  Most formation scenarios of globular clusters commence with 
giant molecular clouds (GMCs) undergoing a phase of rapid star formation.
This star formation can be triggered by several processes, e.g.\ a thermal 
instability (Fall \& Rees 1985; Murray \& Lin 1990), a radiative
shock (Kang et al. 1990; Shapiro 1993) or other perturbations
like collisions of clouds (Fujimoto \& Kumai 1997; Lee, Schramm, 
\& Mathews 1995) or galaxy interactions (Ashman \& Zepf 1992).
A common characteristic of all these scenarios is, that globulars are formed 
from smooth gaseous distributions (if we neglect the clumpy structure of the
GMCs for the moment) which are transformed into stars. This assumption 
leads to several difficulties: First, the 
gravitational binding energy of a homogeneous GMC with $10^6 M_{\odot}$
and a radius of 30 pc is about $2 \cdot 10^{51}$ ergs, whereas it decreases 
for a $10^5 M_{\odot}$ cloud of 10 pc to less than $10^{50}$ ergs. Thus, 
already a 
single supernova injects sufficient energy to destroy a small cloud 
completely, and a few OB stars can even disrupt a $10^6 M_{\odot}$ cloud.
Therefore, 
the formation of the globular must have been finished within a few Myrs,
before the first OB stars explode. A second problem is related to the star
formation efficiency (SFE). Assuming that the GMC is in virial equilibrium
prior to the cluster formation and that the newly born stars keep
the velocity of their parent gas packages, the total energy $E$ of the
stellar system can be estimated by $E = (1/2 \eta - \eta^2) \cdot G M^2 / R$
(with the GMC mass $M$, its radius $R$ and the gravitational constant $G$).
Thus, the system is only gravitationally bound (i.e.\ $E$ becomes negative), 
if the SFE $\eta$, which is the mass fraction of the GMC transformed
into stars, is larger than 50\%. Though mass redistribution in a violent
relaxation stage and a detailed treatment of the energy injection can reduce 
the critical level down to 20\% (Goodwin 1997), the required
SFE still exceeds the typical observed values for GMCs by at least one order
of magnitude (Blitz 1993).

  With respect to these problems, an alternative suggestion by 
Brown, Burkert, \& Truran (1991) is very interesting: 
They suggest that cluster formation starts with an OB-association 
undergoing typeII supernova events near the center of a molecular cloud. 
The expanding supernova remnant sweeps up the cloud material, decelerates 
and might almost be stopped by the external pressure of the ambient hot gas. 
Meanwhile the shell breaks into fragments 
and forms stars. If the total energy of this stellar shell is negative, 
the stars will recollapse and form a bound system. For a simplified
spherical configuration Burkert, Brown, \& Truran (1993) 
demonstrated that the binding
energy of an isolated shell always becomes negative {\it independent} of the
SFE, provided the star formation process does not start too early. 
Thus, the SFE efficiency problem is less severe for this scenario.
In case of an homogeneous ambient medium (e.g.\ in the core of a GMC)
Ehlerov\'a et al.\ (1997) demonstrated that fragmentation in an expanding 
shell takes sufficient time to prevent too rapid star formation.
This result holds also for non-homogeneous power-law density profiles, 
if the density distribution in the GMC is not steeper 
than isothermal (Theis et al.\ 1998).

  According to the shell-scenario the dynamics of its last stage, i.e.\ 
the collapse of a thin stellar shell, is studied and compared with the
collapse of homogeneous spheres representing the standard scenarios. 
N-body simulations are 
performed for isolated configurations as well as for collapses within a
galactic tidal field. The main question addressed here is, whether we can 
discern between different formation scenarios by means of their collapse
dynamics. 


\section{The evolution of an isolated shell}

{\bf Initial conditions and numerical scheme.}
   The shell is modelled by a unit mass which is homogeneously distributed
within the radial range [0.9,1.0] giving a shell thickness of 10\% of the
shell's radius $R$. Initially the shell is at rest. The individual
velocities of the equal mass particles are chosen from an isotropic
velocity distribution resulting in 
a virial coefficient $\eta_{\rm vir} \equiv 2T/|W|=0.05$ 
($T$ is the kinetic and $W$ the potential energy.). 
The simulations are performed with $N=100\,000$ 
particles adopting a softening length $\epsilon$ of 0.01.
The equations of motion are integrated with a leap-frog scheme using a
fixed timestep $\Delta t$ of $10^{-3}$. 
This gives an energy conservation of typically 0.1-0.2\% or better 
over the whole 
integration time. The simulations were performed either with a direct 
summation code (using a GRAPE3af board) or a TREE-code.

{\bf Results.}
   The dynamics of the shell shows three stages. During the first
stage ($t < \tau_{\rm ff} (\rho_{\rm sh})$) the shell is slowly contracting and
small inhomogeneities start to grow (cf.\ also upper left diagram in Fig.\ 1;
$\tau_{\rm ff}(\rho) \equiv [3\pi/(32 G \rho)]^{1/2}$ 
is the free-fall time corresponding to the mass density $\rho$). 
Already during this early stage the particles in the shell are strongly mixed 
because of the radially {\it decreasing} (global) free-fall time in the shell. 
In the second
phase ($\tau_{\rm ff} (\rho_{\rm sh}) < t < 1-2 \tau_{\rm ff} 
({\rm shell})$) the inhomogeneities become 
bound clumps. They merge after the shell's free-fall 
time $\tau_{\rm ff} ({\rm shell}) \sim 1.59$ which exceeds the 
free-fall time of the corresponding sphere by 40\%.
Finally, a radially anisotropic triaxial system is formed.

 The final configuration of the shell simulations is characterized
by a more flattened
shape, a mass-loss of only 8\% (i.e.\ a reduction by a factor of 3.5), 
a smaller anisotropy $1-\sigma^2_\theta/\sigma^2_r$, an increased
half-mass radius (by 50\%) and a decreased 90\% Lagrange radius (factor of 6)
compared to the corresponding collapse of a sphere. Thus, 
violent relaxation is less efficient in case of collapsing shells. 

\begin{figure}[tp]
  \plotone{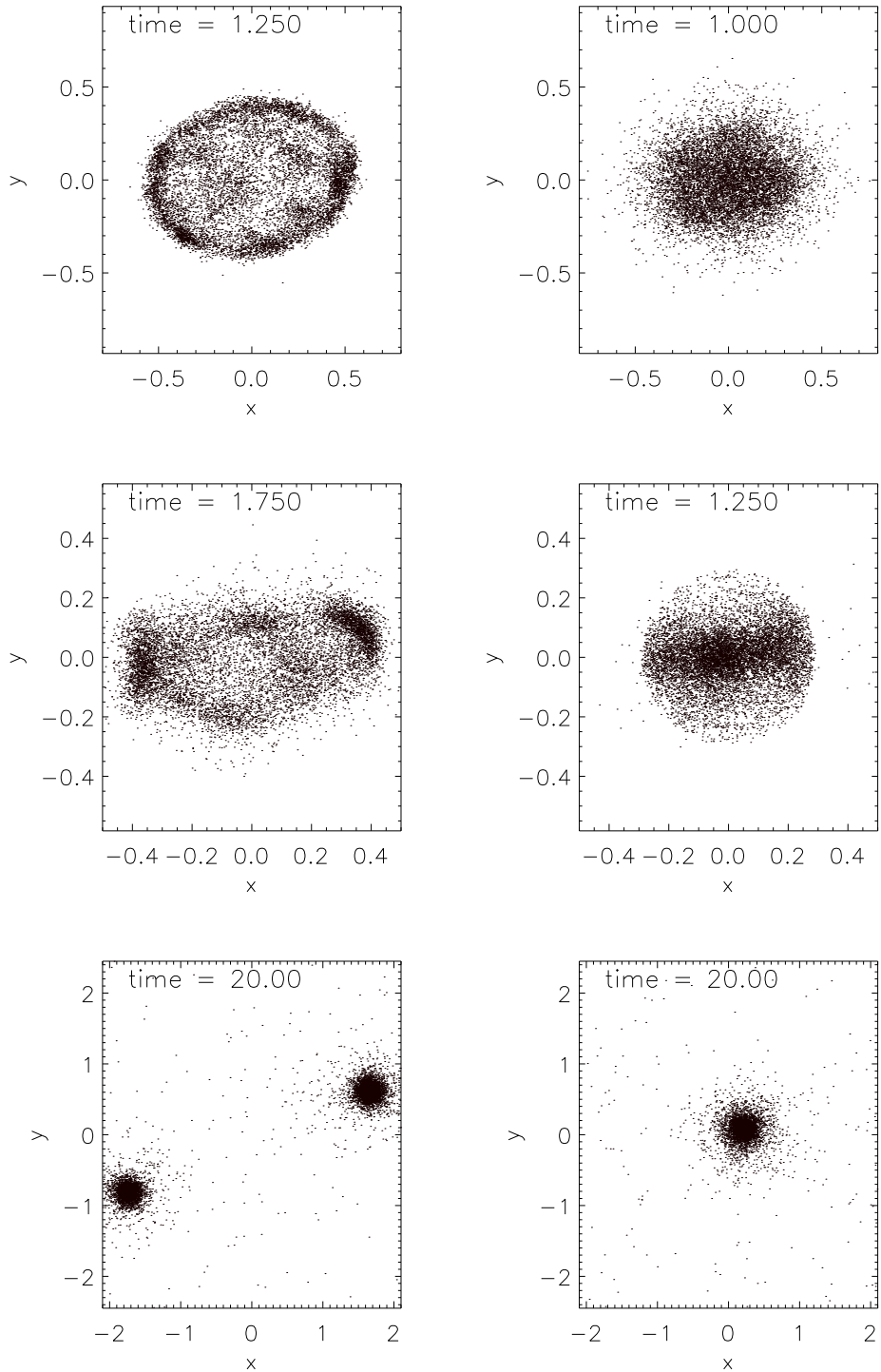}
  \caption{Projections of a collapsing shell (left
   column) and a collapsing sphere (right column) on 
   the orbital plane at different times: a) prior to the collapse (upper row),
   briefly after the collapse (middle row) and in the final stage.
   The units are 7.7 Myrs and 30 pc. Note that the free-fall time for the shell
   is $\sim 1.59$ in the given units, whereas the sphere collapses within
   $t \approx 1.11$.}
\end{figure}


\section{The evolution in a galactic tidal field}

   A realistic model should also include the galactic
tidal field. Therefore, the gravitational force of a static 
isothermal halo with a circular velocity of 220 kms$^{-1}$ 
is added to the force derived from the 
self-gravity of the N-body configurations. In all
following models the galactic orbits have an apogalacticon of 5 kpc
and the clusters start with a mass of $10^5 M_{\odot}$ and a size of 
30 pc. At apogalacticon this corresponds to a tidal radius of
42 pc, i.e.\ systems on circular orbits are not expected to be
tidally disrupted.

{\bf Circular Orbit.}
Though circular orbits are not
very realistic for globular clusters, they keep the tidal field
almost time-independent which allows a more direct investigation of the
influence of the host galaxy. 
Fig.\ 1 shows the evolution of the system projected onto the orbital
plane (and normalized to the center of mass of the particles in the
collapsing system) for both, a shell and a sphere. 
Similarly to the isolated case, we find a strong fragmentation 
prior to the collapse of the whole system in case of the shell, 
whereas almost no substructure is seen during the collapse of the sphere. 
Lateron, the sphere forms a single bound
object which shows immediately after the collapse some elongation caused by 
the tidal field. After one revolution around the galactic center, however, 
the system is almost spherical. In case of the shell, the
influence of the tidal field is already obvious in the distortion
prior to the collapse. The delayed collapse, the fragmentation and also 
the reduced violent relaxation leads to a less dense system which is much
stronger affected by the galactic tidal field: The system does not merge into
a single object, but into two (twin) stellar systems of almost identical 
total mass and density distribution. However, in their kinematical 
properties both objects differ: one of the clusters
shows a counter-rotating core which is neither found in the second
'twin' nor in the product of the collapsing sphere.

\begin{figure}
  \plotone{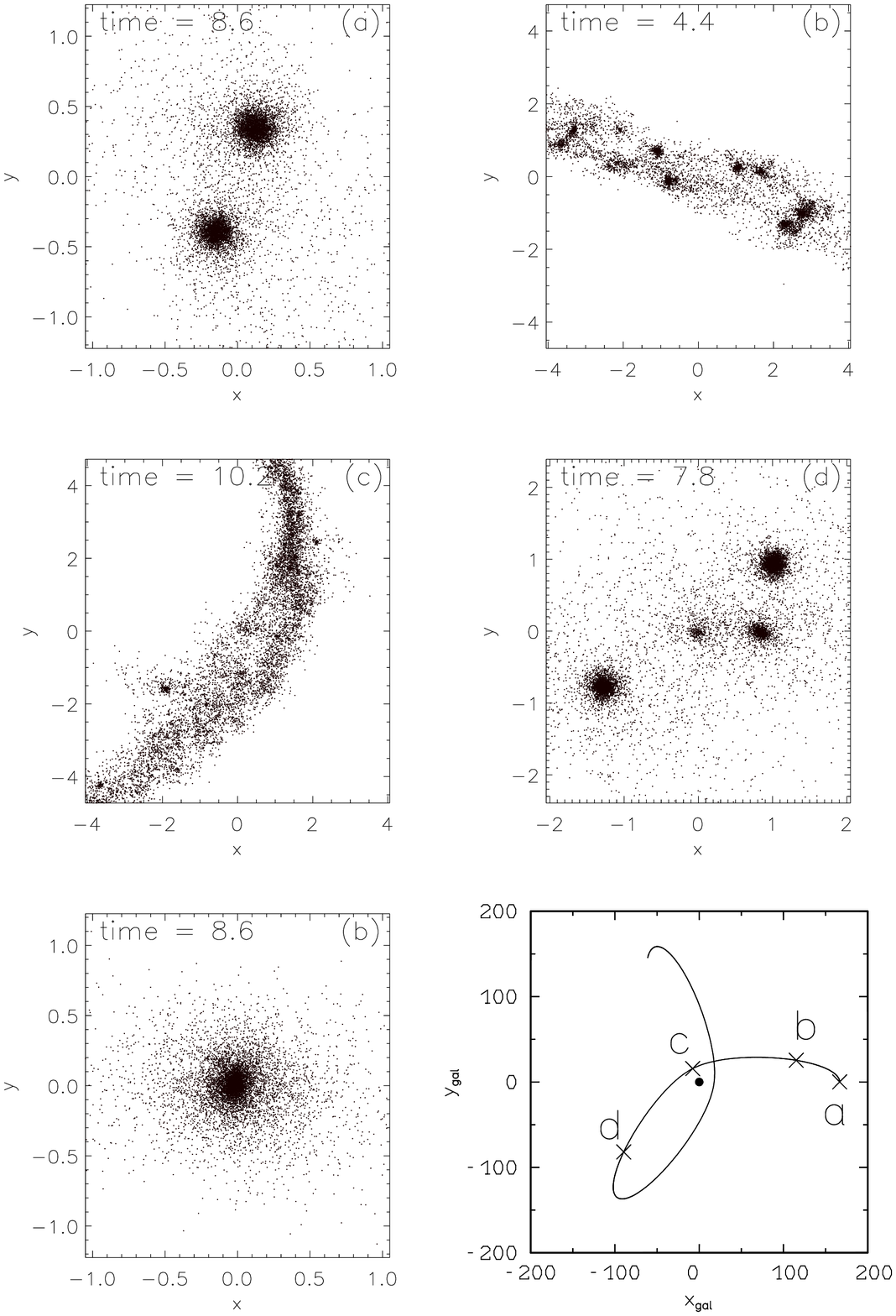}
  \caption{Evolution of shells starting
  to collapse at different phases on an elliptic orbit (upper two rows). 
  The starting points are shown in the lower right panel.
  The lower left diagram displays the final state of an initial sphere which  
  starts at location b. Units and tidal field are identical to Fig.\ 1.}
\end{figure}

  {\bf Eccentric Orbit.} On eccentric orbits the evolution 
additionally depends on the initial phase
(Fig.\ 2). A collapse of a shell starting at apogalacticon ends up again in
a twin cluster (a in Fig.\ 2). 
If the phase is chosen such that maximum compression is 
reached at perigalacticon (b), the system breaks up into many small
systems, but no large cluster is formed. When the
collapse starts at perigalacticon even the formation of these small
clusters is prevented (c). If the collapse starts well after perigalacticon
a multiple system of clusters (with two larger ones) is built up (d).
The final fate of the twins is less clear: the simulations show preferably
surviving twins which separate after a few orbital revolutions to distances
of several 100 pc up to kpc scale or more. 
However, also merging twins have been found.
The low number statistics of the simulations performed so far does not allow 
to give more accurate rates here.
Contrary to the shells the spheres do not form any multiple clusters: either
the systems are completely destroyed or single bound objects are formed.


\section{Summary}

  In case of isolated evolution the simulations 
show several systematic differences between shells and spheres:
The collapse of the shell is delayed, the resulting configurations are 
flatter and (more) triaxial, the mass-loss
is strongly reduced and the system is less dense. All of these effects
suggest that {\it violent relaxation} is less efficient for collapsing
shells. Another difference is the strongly pronounced fragmentation of the
shell which is not seen during the collapse of the sphere. 
Unfortunately, all of these features seem to be very
difficult to be used for an observational discrimination between the 
different formation scenarios. E.g.\ the easy accessible flattening of globular
clusters is not only strongly influenced by the formation scenario, but
also by the secular evolution, the 
galactic tidal field, and by initial conditions like the 
mass distribution in the parent GMC and the virial ratio
of the collapsing stellar system.

  Much more promising are the simulations including a galactic tidal field:
They demonstrated that shells tend to form multiple stellar systems, preferably
twin (but not binary) clusters. In case these twins are not destroyed nor 
merged, one should be able to find pairs of globular clusters which are 
characterized by the properties of their common birth place, i.e.\ 
identical metallicities and
evolutionary stage or similar orbital characteristics. Another
observational feature is the rotation profile of globulars. The merging
of fragments in the shell scenario would give a natural explanation
for counter-rotation in stellar clusters similar to the counter-rotation
found in N-body simulations of merging galaxies.

  A set of movies is available at 
http://www.astrophysik.uni-kiel.de/pershome/theis



\vspace*{3cm}

\subsubsection{Ivan King:}
This is fascinating material. Two Questions: 
1) How much anisotropy is there in your aspherical collapse products?
2) Can you make your animations available, perhaps as mpegs in your anonymous
ftp?
\subsubsection{Christian Theis:}
1) The velocity distributions of both, spheres and shells, are isotropic in 
the central regions. In the outer region
a radial anisotropy (defined here as $1-\sigma^2_\theta/\sigma^2_r$) 
evolves reaching about 0.8 in case of the shell. It is systematically 
below the anisotropy of the sphere by about 0.15.
2) The animations are available via my home-page
http://www.astrophysik.uni-kiel.de/pershome/theis

\subsubsection{Christian Boily:}
1) Why is there no ROI developing in the spherical collapse you showed?
2) Presumably fragmentation here depends on $\sqrt{N}$ noise. What were the
numbers involved here?

\subsubsection{Christian Theis:}

1) The initial virial coefficient here is close to, but exceeding the limit
for the onset of the radial
orbit instability for my simulations. Starting from a lower 
virial ratio of e.g.\ 0.02 shows the ROI for the collapsing spheres as 
expected.\\
2) The fragmentation depends partly on the number of particles and 
and also on the initial setup of the particle configuration 
(here a random realization, i.e.\ white noise). The total number of
particles was $10^5$. However, the most important parameters for the 
onset of fragmentation are the velocity dispersion in the shell and its
thickness. E.g.\ a thickness of 50\% or an initial virial ratio of
0.2 strongly suppresses fragmentation.

\end{document}